\documentstyle[11pt,newpasp,twoside,epsf]{article}
\begin{document}

\title{Stellar Mass-to-Light Ratios and the Tully-Fisher relation}
\author{Roelof S.~de Jong\altaffilmark{1} and Eric F.~Bell}
\affil{Steward Observatory, 933 N. Cherry Ave., Tucson, AZ 85721, U.S.A.}
\altaffiltext{1}{Hubble fellow}

\begin{abstract}
We use spiral galaxy evolution models to argue that there
are substantial variations in stellar mass-to-light ratio ($M/L$)
within and among galaxies.  Our models show a strong correlation
between stellar $M/L$ and galaxy color. 
We compare the colors and maximum-disk $M/L$ values of a sample
of galaxies to the model color-$M/L$ relation, finding that a Salpeter
IMF is too massive but that an IMF with fewer low mass stars 
fits the observations well.
%Comparing the colors and
%maximum-disk $M/L$ values of a sample of galaxies to the model
%color-$M/L$ relation, we find that a Salpeter IMF is too massive, but
%that a reduced low mass star IMF follows the lower envelope in the
%observed color versus maximum-disk $M/L$ distribution. 
Applying our
color-$M/L$ relation to the Tully-Fisher (TF) relation, we find a
stellar mass TF-relation that is independent of originating
passband. Adding the H{\small I} gas mass, we find that the maximum slope
of the baryonic TF-relation is 3.5.
\end{abstract}

\section{Galaxy Evolution Models and Mass-to-Light Ratios}

We have used the galaxy evolution models described by Bell \& Bower
(2000) to investigate stellar $M/L$ ratios of galaxies. These models
were tuned to fit the observed trends between the colors and the
structural parameters of spiral galaxies (Bell \& de Jong 2000). Using a
local gas density dependent star formation law, the photometric
evolution is calculated, taking chemical evolution into account. As well as
a closed box model we have models with gas infall and
outflow, mass dependent formation epochs and star bursts. All models
show large variations in $M/L$, amounting from a factor 8 in $B$ to 2
in $K$, but in all models we find a strong correlation between $M/L$
and optical color (e.g.~Fig.\,1a). The slope of the color-$M/L$
relation is very robust against the particular stellar population
synthesis model and against the exact details of the galaxy
evolution model. The main uncertainty in the correlation is the
zero-point, which is determined by the assumed IMF; most notably by the
relative amount of low mass stars, which contribute to the mass but not
to the luminosity and color.

\begin{figure}
\hbox{%
\epsfysize=4.33cm\epsfbox[184 313 392 519]{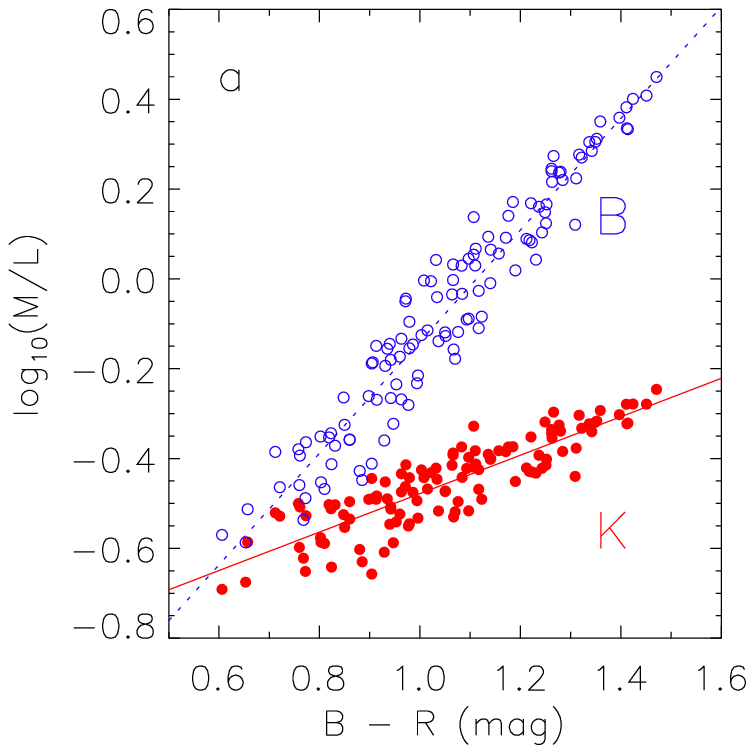}\ 
\epsfysize=4.33cm\epsfbox[184 370 392 576]{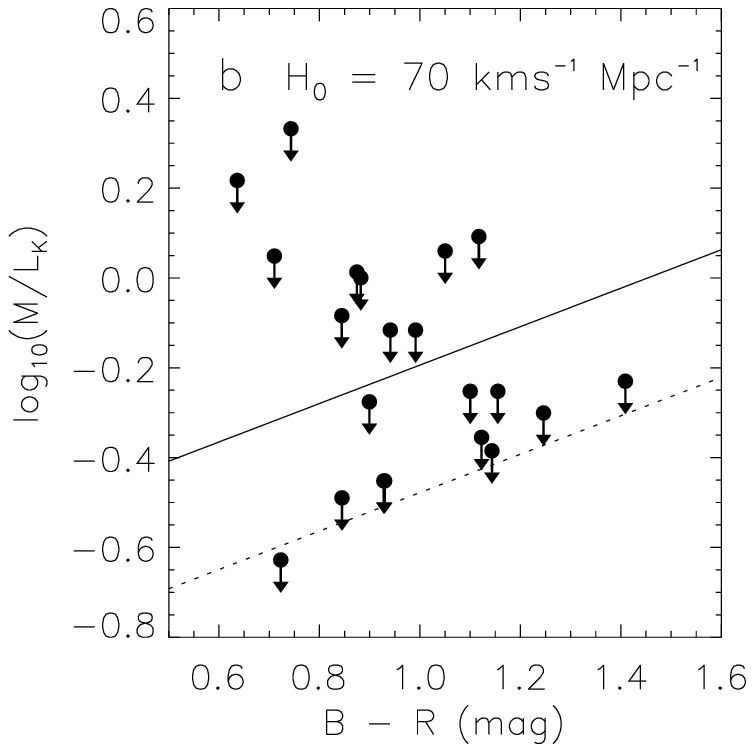}\ 
\epsfysize=4.33cm\epsfbox[184 370 392 576]{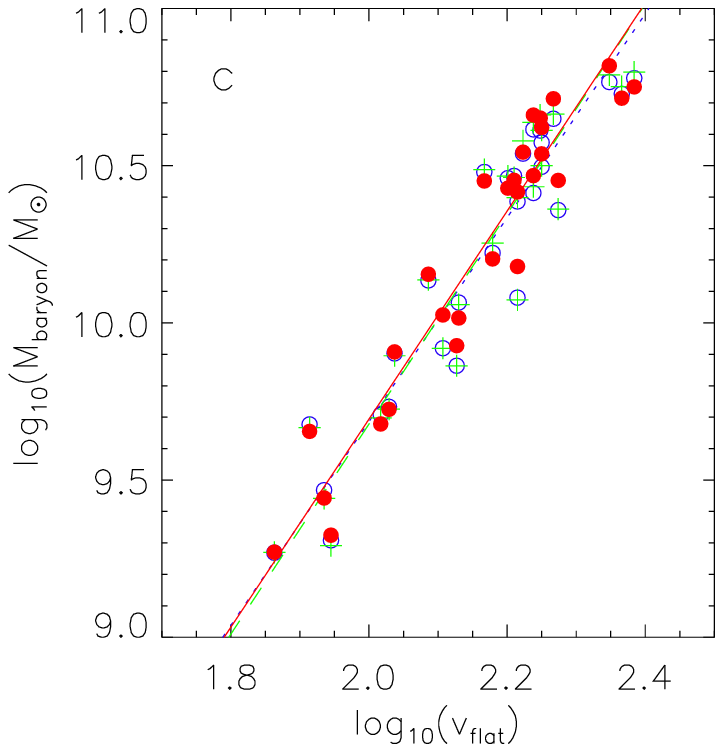}
}
\caption{{\bf a)} $M/L_B$ and $M/L_K$ versus $B$--$R$ for our
mass dependent formation epoch model with star bursts. {\bf b)}
maximum-disk $M/L_K$ versus $B$--$R$ of the Verheijen (1997)
sample. Solid line is a fit to our best Salpeter IMF model, dotted
line modified Salpeter IMF. {\bf c)} Baryonic TF-relations derived
from $B$ ($\circ$), $I$ (+) and $K$ ($\bullet$) observations.}
\end{figure}

%\section{Mass-to-Light Ratio Upper Limits}

A constraint on the color-$M/L$ correlation zero-point can be obtained
from galaxy rotation curves. The stellar disk in a galaxy cannot be
more massive than allowed by its rotation curve, resulting in a
maximum-disk $M/L$. In Fig.\,1b we show the maximum-disk $M/L$ values
versus the extinction corrected $B$--$R$ color for the Verheijen
(1998) galaxy sample. These $M/L$ values are truly upper limits: any
mass not accounted for in the rotation curve decompositions will push
the stellar $M/L$ even lower.  The solid line in Fig.\,1b shows the
fit to the color-$M/L$ relation for our best model using a standard
Salpeter IMF. Clearly this model over-predicts the maximum
allowed mass for many galaxies, as many galaxies lie below the
line. Using a Salpeter IMF with a flat slope below 0.6\,$M_\odot$ as
suggested by recent observations results in the dotted line which is
consistent with the observations (for $H_0 =
70$\,km\,s$^{-1}$\,Mpc$^{-1}$ or $D_{\,\rm Ursa Major} = 20$ Mpc). 
%and actually follows the lower
%envelope of the color versus maximum disk $M/L$ distribution.

\section{Tully-Fisher Relations}

Observed TF-relations are known to have a passband dependence, both in
slope and in zero-point.  Applying Tully et al.~(1998) extinction
corrections and our color-$M/L$ correlations, we can calculate stellar
mass TF-relations from the observed TF-relations. We find that the
stellar mass TF-relations derived from the different passbands are
equal to within the uncertainties. By adding in the H{\small I} gas mass
we can calculate the baryonic TF-relations (Fig.\,1c). We find
that the slope of the baryonic TF-relation must be less then
$3.5\pm0.2$, significantly lower than found by McGaugh et al.~(2000),
%mainly due to our exclusion of low luminosity dwarfs with poorly
%determined inclinations and rotation velocities and due to our lower
%$H_0$ value (65 versus 75 km\,s$^{-1}$\,Mpc$^{-1}$).
mainly due to our use of stellar M/Ls consistent with maximum
disk constraints and our exclusion of low luminosity dwarfs with
poorly determined inclinations and rotation velocities.

\acknowledgements

Support for RSdJ was provided by NASA through Hubble Fellowship grant
\#HF-01106.01-A from the Space Telescope Science Institute, which is
operated by the AURA, Inc., under NASA contract NAS5-26555.  Support
for EFB was provided by NASA LTSA grant NAG5-8426.

\end{document}